\documentclass[preprint]{revtex4}
\usepackage{epsfig}

\begin{document}

\title{KdV solitons in a cold quark gluon plasma}

\author{D.A. Foga\c{c}a\dag\,  F.S. Navarra\dag\ and L.G. Ferreira Filho\ddag\ }
\address{\dag\ Instituto de F\'{\i}sica, Universidade de S\~{a}o Paulo\\
 C.P. 66318,  05315-970 S\~{a}o Paulo, SP, Brazil}
\address{\ddag\ Faculdade de Tecnologia, Universidade do Estado do Rio de Janeiro \\
Via Dutra km 298, CEP 27523-000, Resende, RJ, Brazil}

\begin{abstract}

The relativistic heavy ion  program developed at RHIC and now at LHC motivated a deeper
study of the properties of the quark gluon plasma (QGP) and, in particular, the study of 
perturbations in this kind of plasma. We are interested on the time evolution of 
perturbations in the baryon and energy densities. If a localized pulse in baryon density  
could propagate throughout  the QGP for long distances preserving its shape and without 
loosing localization, this could have interesting consequences for relativistic heavy ion 
physics and for astrophysics. A mathematical way to proove that this can happen is to derive 
(under certain conditions) from the hydrodynamical equations of the QGP a  Korteveg-de 
Vries (KdV) equation. The solution of this equation  describes the propagation of a KdV 
soliton. The derivation of the KdV equation depends crucially on the equation of state 
(EOS) of the QGP. The use of the simple MIT bag model EOS does not lead to KdV solitons. 
Recently we have developed an EOS for the  QGP which includes both perturbative and 
non-perturbative corrections  to the MIT one and is still simple enough to allow for 
analitycal manipulations. With this EOS we were able to derive a KdV equation for the 
cold QGP. 
 
\end{abstract} 

%\pacs{PACS Numbers~ :~ 12.38.Lg, 12.40.Yx, 12.39.Mk}   
\maketitle

%%%%%%%%%%%%%%%%%%%%%%%%%%%%%%%%%%%%%%%%%%%%%%%%%%%%%%%%%%%%%%%%%%%%%%%%%
% Beginning of the paper
%%%%%%%%%%%%%%%%%%%%%%%%%%%%%%%%%%%%%%%%%%%%%%%%%%%%%%%%%%%%%%%%%%%%%%%%%

%%%%%%%%%%%%%%%%%%%  Introduction %%%%%%%%%%%%%%%%%%%%%%%%%%%%%%%%%%%%%%%

\vspace{1cm}
\section{Introduction}

Korteweg - de Vries solitons are  very interesting non-linear waves, which may exist 
in many types of fluids from ordinary water to astrophysical plasmas \cite{drazin}.  
In the last years 
we have started to produce a new kind of fluid in laboratory: the quark gluon plasma (QGP). 
This is a state where quarks and gluons, usually confined in the interior of baryons 
(such as the proton) and mesons, are free to travel longer distances. With the beginning  
of the LHC era, we have means to study larger and longer living samples of QGP and even the 
propagation of perturbations in this new medium. In this context a natural question is: 
can we have KdV solitons in the QCD plasma? In this work we give an answer to this question.
 
Before the QGP there were other fluids made of strongly interacting hadronic matter and the 
existence of KdV solitons in these fluids was already investigated. The first works 
on the subject were published in \cite{frsw}, where the authors considered the propagation of
baryon density pulses in proton-nucleus collisions at intermediate energies. In this scenario 
the incoming proton would be absorbed by the nuclear fluid generating a KdV soliton, which, 
traversing the whole nucleus without distortion, would escape from the target as a proton and 
would simulate an unexpected transparency. In \cite{frsw} the existence of the KdV soliton 
relied solely on the equation of state (EOS), which had no deep  
justification. In \cite{fn1} we 
have reconsidered the problem, introducing an equation of state derived from relativistic mean
field models of nuclear matter.  We concluded that the homogeneous meson field approximation 
was 
too strong and would exclude the existence of KdV solitons. We could also trace back the 
derivative terms in the energy density to derivative couplings between the nucleon and the 
vector meson. In \cite{fn2} we extended our analysis 
to relativistic hydrodynamics and in \cite{fn3} to spherical and cylindrical geometries. In 
\cite{fn4} we considered hadronic matter at finite temperature and studied the effects of 
temperature on the KdV soliton. In \cite{fn5} we started the study of perturbations in the 
QGP at zero and finite temperature. The  conclusion found in that work was that the existence 
of KdV solitons in a QGP depends on details of the EOS and with a simple MIT bag model EOS 
there is no KdV soliton! A further study of the equation of state, carried out in \cite{fn6}, 
showed that if non-perturbative effects are included in the EOS through gluon condensates, 
then new terms 
appear in the expression of the energy density and pressure and in the present work we show 
how these new terms lead to a KdV equation, after the proper treatment of the hydrodynamical 
equations. 

In the next section we briefly  review the equations of one-dimensional 
relativistic fluid dynamics. 
In section III we introduce the equation of state, in section IV we derive the KdV equation 
and in section V we present a  numerical analysis of the obtained equation.  

\section{Relativistic Fluid Dynamics}

Relativistic hydrodynamics is well presented in the textbooks  \cite{wein,land}. 
The relativistic version of the Euler 
equation \cite{wein,land,fn5} is given by:
\begin{equation}
{\frac{\partial {\vec{v}}}{\partial t}}+(\vec{v} \cdot \vec{\nabla}) \, \vec{v}=
-{\frac{1}{(\varepsilon + p)\gamma^{2}}}
\bigg({\vec{\nabla} p + \vec{v} \,  {\frac{\partial p}{\partial t}}}\bigg)
\label{eul}
\end{equation}
where $\vec{v}$, $\varepsilon$, $p$ and $\gamma$ are the velocity, energy density, 
pressure and the Lorentz factor respectively. We employ the natural units $c=1$ and 
$\hbar=1$. Space and time coordinates will be in $fm$ ($1 fm = 10^{-15} m$).
The relativistic version of the continuity equation for the baryon 
density is 
\cite{wein}:
\begin{equation}
\partial_{\nu}{j_{B}}^{\nu}=0
\label{conucleon}
\end{equation}
Since ${j_{B}}^{\nu}=u^{\nu} \rho_{B}$ the above equation can be rewritten as \cite{fn5}:
\begin{equation}
{\frac{\partial \rho_{B}}{\partial t}}+\gamma^{2} \vec{v}  \, 
\rho_{B}\Bigg({\frac{\partial \vec{v} }
{\partial t}}+ \vec{v} \cdot \vec{\nabla} \vec{v}\Bigg)+\vec{\nabla} \cdot 
(\rho_{B} \, \vec{v})=0
\label{rhobcons}
\end{equation}
where $\rho_{B}$ is the baryon density. In the one dimensional 
Cartesian relativistic fluid dynamics the velocity field is written as 
$\vec{v}=v(x,t) \, \hat{x}$ where $\hat{x}$ is the unit vector in the  $x$ direction.  
Equations (\ref{eul}) and (\ref{rhobcons}) can be rewritten in the simple form:
\begin{equation}
{\frac{\partial v}{\partial t}}+v{\frac{\partial v}{\partial x}}=
{\frac{(v^{2}-1)}{(\varepsilon + p)}}
\bigg({\frac{\partial p}{\partial x}}+v{\frac{\partial p}{\partial t}} \bigg)
\label{eulu}
\end{equation}
and
\begin{equation}
v\rho_{B}\bigg({\frac{\partial v}{\partial t}}+v{\frac{\partial v}{\partial x}}\bigg)+
(1-v^{2})
\bigg({\frac{\partial \rho_{B}}{\partial t}}+\rho_{B}{\frac{\partial v}{\partial x}}+
v{\frac{\partial \rho_{B}}{\partial x}} \bigg)=0
\label{rhobconsu}
\end{equation}

\section{The QGP Equation of State} 

In what follows we  present the  mean field treatment of  QCD 
developed in   \cite{fn6} (for previous works on the subject see 
\cite{shakin,shakinn}) and go beyond the homogeneous field approximation, including 
the terms with gradients.

The Lagrangian density of QCD is given by:
\begin{equation}
{\mathcal{L}}_{QCD}=-{\frac{1}{4}}F^{a}_{\mu\nu}F^{a\mu\nu}
+\sum_{q=1}^{N_{f}}\bar{\psi}^{q}_{i}\Big[i\gamma^{\mu}(\delta_{ij}\partial_{\mu}-
igT^{a}_{ij}G_{\mu}^{a})
- \delta_{ij} m_q \Big]\psi^{q}_{j}
\label{lqcdu}
\end{equation}
where
\begin{equation}
F^{a\mu\nu}=\partial^{\mu}G^{a\nu}-\partial^{\nu}G^{a\mu}+gf^{abc}G^{b\mu}G^{c\nu}
\label{efe}
\end{equation}
The summation on $q$ runs over all quark flavors, 
$m_q$ is the mass of the quark of flavor $q$,  
$i$ and $j$ are the color indices of the quarks, 
$T^{a}$ are the SU(3) generators and $f^{abc}$ are the SU(3) antisymmetric 
structure constants.  For simplicity we will consider massless quarks, i.e. $m_q=0$.  
Moreover, we will drop the summation and consider only 
one flavor. At the end of our calculation the number of flavors will be recovered.  
Following \cite{shakin, shakinn}, we shall  write the gluon field as:
\begin{equation}
G^{a\mu}={A}^{a\mu}+{\alpha}^{a\mu}
\label{amd}
\end{equation}
where ${A}^{a\mu}$  and ${\alpha}^{a\mu}$ are the low (``soft'') and high (``hard'')  momentum
components of the gluon field respectively. We will assume that ${A}^{a\mu}$ 
represents the soft modes which populate the vacuum and the terms containing  ${A}^{a\mu}$ will
be replaced  by their expectation values   $\langle {A}^{a\mu} \rangle$, 
 $\langle {A}^{a\mu} {A}^{a}_{\mu}  \rangle$, etc...in the plasma. ${\alpha}^{a\mu}$  
represents the modes for which the running coupling constant is small. 

%Inserting (\ref{amd}) into (\ref{efe}) we obtain:
%$$
%F^{a\mu\nu}= (\partial^{\mu}A^{a\nu}-\partial^{\nu}A^{a\mu}+ g f^{abc}A^{b\mu}A^{c\nu})
%+(\partial^{\mu}{\alpha}^{a\nu}-\partial^{\nu}{\alpha}^{a\mu} + 
%g f^{abc}{\alpha}^{b\mu}{\alpha}^{c\nu}) 
%$$
%\begin{equation}
%+ g f^{abc} {A}^{b\mu} {\alpha}^{c\nu} 
%+ g f^{abc} {\alpha}^{b\mu} {A}^{c\nu} 
%\label{efesplit}
%\end{equation}

In a cold quark gluon plasma the density is much larger than the ordinary nuclear matter 
density. These high densities imply a very large number of sources of the gluon field. 
Assuming that the coupling constant is not very small, the existence of  intense 
sources implies that  the bosonic fields tend to have large occupation numbers at all 
energy levels, and therefore they can be treated as classical fields. 
This is the famous approximation for 
bosonic fields used in relativistic mean field models of nuclear matter \cite{serot}. 
It has been applied to QCD in the past  and amounts to assume that the 
``hard'' gluon field, ${\alpha}_{\mu}^{a}$,  is simply a  function of the coordinates:
\begin{equation}
{\alpha}_{\mu}^{a}(\vec{x},t)=\delta_{\mu 0} \, {\alpha}_{0}^{a}(\vec{x},t)
\label{watype}
\end{equation}
with  $\partial_{\nu}{\alpha}^{a}_{\mu}\neq 0$. This space and time dependence goes 
beyond the standard mean field approximation \cite{serot},  where  
${\alpha}_{\mu}^{a}$ is constant in space and time  and 
consequently $\partial_{\nu}{\alpha}^{a}_{\mu}=0$.
We keep  assuming, as in \cite{fn6},  that the soft gluon field ${A}^{a\mu}$ is 
independent of position and time and thus $\partial^{\nu}{A}^{a\mu}=0$ .  
Following the same steps introduced in
\cite{fn6} we obtain the following effective Lagrangian:
\begin{equation}
\mathcal{L}_{0}=-{\frac{1}{2}}{\alpha}^{a}_{0}\big({\vec{\nabla}}^{2}{\alpha}^{a}_{0}\big)
+{\frac{{m_{G}}^{2}}{2}}{\alpha}^{a}_{0}{\alpha}^{a}_{0}-\mathcal{B}_{QCD}
+ \bar{\psi}_{i}\Big(i\delta_{ij}\gamma^{\mu}\partial_{\mu}+g\gamma^{0}T^{a}_{ij}
{\alpha}^{a}_{0}  \Big)\psi_{j}
\label{mfqcdf}
\end{equation}
where  $m_G$ is the  dynamical  mass of the hard gluon $\alpha$ generated by its interaction 
with the soft gluons $A^{a \, \mu}$ from the vacuum and it is related to the dimension two  
$\langle A^2 \rangle$ gluon condensate. The constant $\mathcal{B}_{QCD}$ is related to the 
dimension four gluon condensate $\langle F^2 \rangle$ (see \cite{fn6} for details).

The effective Lagrangian (\ref{mfqcdf}) is quite similar to the one  
obtained in \cite{fn6} and the only difference is the first term, which is new and comes 
from the gradients. The equations of motion \cite{fn4} are given by:
\begin{equation}
{\frac{\partial \mathcal{L}}{\partial \eta_{i}}}
-\partial_{\mu}{\frac{\partial \mathcal{L}}{\partial(\partial_{\mu} \eta_{i})}}
+\partial_{\nu}\partial_{\mu}\bigg[{\frac{\partial \mathcal{L}}
{\partial(\partial_{\mu}\partial_{\nu} \eta_{i})}}\bigg]=0
\label{eulerlagra}
\end{equation}
Inserting (\ref{mfqcdf}) into (\ref{eulerlagra}) with
$\eta_{1}={\alpha}^{a}_{0}(\vec{x},t)$ and $\eta_{2}=\bar{\psi}(\vec{x},t)$ we find:
\begin{equation}
-{\vec{\nabla}}^{2}{\alpha}^{a}_{0} + {m_{G}}^{2}{\alpha}^{a}_{0}=-g\rho^{a}
\label{azeaem}
\end{equation}
\begin{equation}
\Big(i\gamma^{\mu}\partial_{\mu}+g\gamma^{0}T^{a}{\alpha}^{a}_{0} \Big)\psi=0
\label{psiem}
\end{equation}
where  $\rho^{a}$ is the temporal component of the  color vector current  given by 
$j^{a\nu}=\bar{\psi}_{i}\gamma^{\nu}T^{a}_{ij}\psi_{j}$ .
The energy-momentum tensor reads \cite{fn4}:
\begin{equation}
T^{\mu \nu}={\frac{\partial \mathcal{L}}{\partial(\partial_{\mu}\eta_{i})}}
(\partial^{\nu}\eta_{i})
-g^{\mu \nu}\mathcal{L} - \bigg[\partial_{\beta}
{\frac{\partial \mathcal{L}}{\partial (\partial_{\mu}\partial_{\beta} 
\eta_{i})}}\bigg] (\partial^{\nu}\eta_{i})+
{\frac{\partial \mathcal{L}}{\partial (\partial_{\mu}\partial_{\beta}
 \eta_{i})}}(\partial_{\beta}\partial^{\nu}\eta_{i})
\label{tensorem}
\end{equation}
From the above expression we can obtain  the energy density ($\varepsilon=<T_{00}>$)
which turns out to be \cite{fn6}:
%\begin{equation}
%\varepsilon={\frac{1}{2}}{\alpha}^{a}_{0}\big({\vec{\nabla}}^{2}{\alpha}^{a}_{0}\big)
%-{\frac{{m_{G}}^{2}}{2}}{\alpha}^{a}_{0}{\alpha}^{a}_{0}+b{\phi_{0}}^{4}+
%i\bar{\psi}\gamma^{0}(\partial_{0}\psi)
%\label{epsilonnn}
%\end{equation}
%Rewritting the last term of the above expression \cite{fn6} we arrive at:
\begin{equation}
\varepsilon={\frac{1}{2}}{\alpha}^{a}_{0}\big({\vec{\nabla}}^{2}{\alpha}^{a}_{0}\big)
-{\frac{{m_{G}}^{2}}{2}}{\alpha}^{a}_{0}{\alpha}^{a}_{0}+ \mathcal{B}_{QCD}
-g\rho^{a}{\alpha}^{a}_{0}
+3{\frac{\gamma_{Q}}{2{\pi}^{2}}} \int_{0}^{k_{F}}
dk \hspace{0.1cm} k^{2}\sqrt{{\vec{k}}^{2}+m^{2}}
\label{epsilonnnn}
\end{equation}
where  $\gamma_{Q}$ is the quark degeneracy factor 
$\gamma_{Q} = 2 (\mbox{spin}) \times 3 (\mbox{flavor}) $. The sum over all the color 
states was 
already performed and resulted in the pre-factor $3$ in the expression above. 
$k_{F}$ is the Fermi momentum defined by the quark number density $\rho$:
\begin{equation}
\rho=\langle N | \psi^{\dagger}_{i} \psi_{i} | N \rangle= {\frac{3}{V}}\sum_{\vec{k},\lambda}
\langle N | N \rangle=3{\frac{\gamma_{Q}}{(2\pi)^{3}}}
\int d^{3}k
=3{\frac{\gamma_{Q}}{2{\pi}^{2}}} \int_{0}^{k_{F}} dk \hspace{0.1cm} k^{2}  
= {\frac{\gamma_{Q}}{2{\pi}^{2}}}{{k_{F}}}^{3}
\label{kf}
\end{equation}
In the above expression $|N \rangle$ denotes a state with N quarks.
In a first approximation the field ${\alpha}^{a}_{0}$ may be estimated  from (\ref{azeaem}).
Neglecting  the  derivative term  ${\vec{\nabla^{2}}}{\alpha}^{a}_{0}$ of (\ref{azeaem}) we have:
\cite{fn4}:
\begin{equation}
{\alpha}^{a}_{0}\cong -{\frac{g}{{m_{G}}^{2}}}\rho^{a}
\label{azeropest}
\end{equation}
Inserting (\ref{azeropest}) in the first term of (\ref{azeaem}) and then solving
it for ${\alpha}^{a}_{0}$ we find:
\begin{equation}
{\alpha}^{a}_{0}=-{\frac{g}{{m_{G}}^{2}}}\rho^{a}-{\frac{g}
{{m_{G}}^{4}}}{\vec{\nabla}}^{2}\rho^{a}
\label{azerototale}
\end{equation}
We can write the color charge density $\rho^{a}$ in terms of the quark number density $\rho$ 
through:
\begin{equation}
\rho^{a}\rho^{a}=3\rho^{2}
\label{roqaa}
\end{equation}
Analogously we have
\begin{equation}
\rho^{a}\vec{\nabla}^{2}\rho^{a}=3\rho\vec{\nabla}^{2}\rho \hspace{0.1cm}, \hspace{0.3cm}
\rho^{a}\vec{\nabla}^{4}\rho^{a}=3\rho\vec{\nabla}^{4}\rho \hspace{0.1cm} \hspace{0.3cm} 
\label{roqaaa}
\end{equation}
Inserting (\ref{azerototale}), (\ref{roqaa}) and (\ref{roqaaa}) into (\ref{epsilonnnn}),
performing the momentum integral and using the baryon density, 
which is $\rho_{B}={\frac{1}{3}}\rho$, we arrive at the final expression for the energy 
density in one spatial dimension:
$$
\varepsilon=\bigg({\frac{27g^{2}}{2{m_{G}}^{2}}}\bigg)  {\rho_{B}}^{2}
+\bigg({\frac{27g^{2}}{2{m_{G}}^{4}}}\bigg) 
\rho_{B}  {\frac{\partial^{2}\rho_{B}}{\partial x^{2}}}
+\bigg({\frac{27g^{2}}{2{m_{G}}^{6}}}\bigg)  \rho_{B}  
{\frac{\partial^{4}\rho_{B}}{\partial x^{4}}}
+\bigg({\frac{27g^{2}}{2{m_{G}}^{8}}}\bigg)  
{\frac{\partial^{2}\rho_{B}}{\partial x^{2}}} \  {\frac{\partial^{4}\rho_{B}}{\partial x^{4}}}
$$
%$$
%+\bigg({\frac{27g^{2}}{4{m_{G}}^{6}}}\bigg) \ {\frac{\partial^{2}\rho_{B}}
%{\partial x^{2}}}\ {\frac{\partial^{2}\rho_{B}}{\partial t^{2}}}
%-\bigg({\frac{27g^{2}}{4{m_{G}}^{6}}}\bigg) \ \rho_{B} \ {\frac{\partial^{2}}
%{\partial x^{2}}}\Bigg({\frac{\partial^{2}\rho_{B}}{\partial t^{2}}}\Bigg)
%-\bigg({\frac{27g^{2}}{4{m_{G}}^{8}}}\bigg) \ {\frac{\partial^{2}\rho_{B}}
%{\partial x^{2}}} \ 
%{\frac{\partial^{2}}{\partial x^{2}}} \Bigg({\frac{\partial^{2}\rho_{B}}
%{\partial t^{2}}}\Bigg)
%$$
%$$
%-\bigg({\frac{27g^{2}}{4{m_{G}}^{8}}}\bigg) \ {\frac{\partial^{4}\rho_{B}}
%{\partial x^{4}}} \ {\frac{\partial^{2}\rho_{B}}{\partial t^{2}}}
%+\bigg({\frac{27g^{2}}{8{m_{G}}^{8}}}\bigg) \ {\frac{\partial^{2}\rho_{B}}
%{\partial t^{2}}} \ 
%{\frac{\partial^{2}}{\partial x^{2}}}\Bigg({\frac{\partial^{2}\rho_{B}}{\partial t^{2}}}\Bigg)
%-\bigg({\frac{27g^{2}}{8{m_{G}}^{6}}}\bigg){\frac{\partial^{2}\rho_{B}}
%{\partial t^{2}}} \ {\frac{\partial^{2}\rho_{B}}{\partial t^{2}}}
%$$
\begin{equation}
+\mathcal{B}_{QCD}
+3{\frac{\gamma_{Q}}{2{\pi}^{2}}}{\frac{{k_{F}}^{4}}{4}}
\label{eps}
\end{equation}
The pressure is given by $p={\frac{1}{3}}<T_{ii}>$. Repeating the same steps mentioned 
before we arrive at:
$$
p=\bigg({\frac{27g^{2}}{2{m_{G}}^{2}}}\bigg) \ {\rho_{B}}^{2}+\bigg({\frac{18g^{2}}{{m_{G}}^{4}}}\bigg) \
\rho_{B} {\frac{\partial^{2}\rho_{B}}{\partial x^{2}}}
-\bigg({\frac{9g^{2}}{{m_{G}}^{6}}}\bigg) \rho_{B} {\frac{\partial^{4}\rho_{B}}{\partial x^{4}}}
-\bigg({\frac{9g^{2}}{2{m_{G}}^{4}}}\bigg){\frac{\partial\rho_{B}}{\partial x}}
{\frac{\partial\rho_{B}}{\partial x}}
$$
$$
+\bigg({\frac{9g^{2}}{2{m_{G}}^{6}}}\bigg)
{\frac{\partial^{2}\rho_{B}}{\partial x^{2}}}{\frac{\partial^{2}\rho_{B}}{\partial x^{2}}}
-\bigg({\frac{9g^{2}}{{m_{G}}^{8}}}\bigg) {\frac{\partial^{2}\rho_{B}}{\partial x^{2}}} \
{\frac{\partial^{4}\rho_{B}}{\partial x^{4}}}
-\bigg({\frac{9g^{2}}{2{m_{G}}^{8}}}\bigg) {\frac{\partial^{3}\rho_{B}}{\partial x^{3}}} \
{\frac{\partial^{3}\rho_{B}}{\partial x^{3}}}
-\bigg({\frac{9g^{2}}{{m_{G}}^{6}}}\bigg) {\frac{\partial\rho_{B}}{\partial x}} \
{\frac{\partial^{3}\rho_{B}}{\partial x^{3}}}
$$
\begin{equation}
-\mathcal{B}_{QCD}
+{\frac{\gamma_{Q}}{2{\pi}^{2}}}{\frac{{k_{F}}^{4}}{4}}
\label{pres}
\end{equation}
where now $k_{F}$ defined by (\ref{kf}) is given by $\rho_{B}={k_{F}}^{3}/{\pi}^{2}$ .

\section{The KdV Equation}

We now combine the equations (\ref{eulu}) and (\ref{rhobconsu}) to obtain the KdV equation
which governs the space-time evolution of the perturbation in the baryon density.  
We first write (\ref{eulu}) and (\ref{rhobconsu}) in terms of the dimensionless variables:
\begin{equation}
\hat\rho={\frac{\rho_{B}}{\rho_{0}}} \hspace{0.2cm}, \hspace{0.5cm} \hat v={\frac{v}{c_{s}}}
\label{vadima}
\end{equation}
where $\rho_0$ is an equilibrium (or reference) density, upon which perturbations may be 
gene-rated, and $c_s$ is the speed of sound. 
Next, we introduce the  $\xi$ and $\tau$ ``stretched'' coordinates 
\cite{frsw,davidson}:
\begin{equation}
\xi=\sigma^{1/2}{\frac{(x-{c_{s}}t)}{R}} 
\hspace{0.2cm}, \hspace{0.5cm} 
\tau=\sigma^{3/2}{\frac{{c_{s}}t}{R}} 
\label{streta}       
\end{equation} 
where $\sigma$ is a small expansion parameter, $R$ is a typical size scale of the problem.  
After this change of variables  we  expand   (\ref{vadima}) as:
\begin{equation}
\hat\rho=1+\sigma \rho_{1}+ \sigma^{2} \rho_{2}+ \dots
\label{roexpa}
\end{equation}
\begin{equation}
\hat v=\sigma v_{1}+ \sigma^{2} v_{2}+ \dots
\label{vexpa}
\end{equation} 
Having rewritten  (\ref{eulu}) and (\ref{rhobconsu})  in the $\xi-\tau$ space and having 
expanded them in powers of $\sigma$  up to $\sigma^2$ we organize the two   
equations as series in powers of $\sigma$.  After these steps 
(\ref{eulu}) and (\ref{rhobconsu}) become:
$$
\sigma \Bigg\lbrace -\bigg[\bigg({\frac{27g^{2}\,{\rho_{0}}^{2}}{{m_{G}}^{2}}}\bigg){c_{s}}^{2}
+3\pi^{2/3}{\rho_{0}}^{4/3}{c_{s}}^{2}\bigg]
{\frac{\partial v_{1}}{\partial \xi}}+\bigg[\bigg({\frac{27g^{2}\,{\rho_{0}}^{2}}{{m_{G}}^{2}}}\bigg)
+\pi^{2/3}{\rho_{0}}^{4/3}\bigg]
{\frac{\partial\rho_{1}}{\partial \xi}} \Bigg\rbrace
$$
$$
+\sigma^{2}\Bigg\lbrace \bigg[\bigg({\frac{27g^{2}\,{\rho_{0}}^{2}}{{m_{G}}^{2}}}\bigg)
+\pi^{2/3}{\rho_{0}}^{4/3}\bigg]{\frac{\partial\rho_{2}}{\partial \xi}}
-\bigg[\bigg({\frac{27g^{2}\,{\rho_{0}}^{2}}{{m_{G}}^{2}}}\bigg){c_{s}}^{2}
+3\pi^{2/3}{\rho_{0}}^{4/3}{c_{s}}^{2}\bigg]{\frac{\partial v_{2}}{\partial \xi}}
$$
$$
+\bigg[\bigg({\frac{27g^{2}\,{\rho_{0}}^{2}}{{m_{G}}^{2}}}\bigg){c_{s}}^{2}
+3\pi^{2/3}{\rho_{0}}^{4/3}{c_{s}}^{2}\bigg]\bigg({\frac{\partial v_{1}}{\partial \tau}}
+v_{1}{\frac{\partial v_{1}}{\partial \xi}}\bigg)+
\bigg({\frac{27g^{2}\,{\rho_{0}}^{2}}{{m_{G}}^{2}}}\bigg)\rho_{1}{\frac{\partial \rho_{1}}{\partial \xi}}
+\pi^{2/3}{\rho_{0}}^{4/3}{\frac{\rho_{1}}{3}}{\frac{\partial \rho_{1}}{\partial \xi}}
$$
$$
-\bigg[\bigg({\frac{27g^{2}\,{\rho_{0}}^{2}}{{m_{G}}^{2}}}\bigg)2{c_{s}}^{2}
+4\pi^{2/3}{\rho_{0}}^{4/3}{c_{s}}^{2}\bigg]\rho_{1}{\frac{\partial v_{1}}{\partial \xi}}
-\bigg[\bigg({\frac{27g^{2}\,{\rho_{0}}^{2}}{{m_{G}}^{2}}}\bigg){c_{s}}^{2}
+\pi^{2/3}{\rho_{0}}^{4/3}{c_{s}}^{2}\bigg]v_{1}{\frac{\partial \rho_{1}}{\partial \xi}}
$$
\begin{equation}
+\bigg({\frac{18g^{2}\,{\rho_{0}}^{2}}{{m_{G}}^{4}R^{2}}}\bigg)
{\frac{\partial^{3} \rho_{1}}{\partial \xi^{3}}} \Bigg\rbrace =0
\label{eulerexp}
\end{equation}
and
\begin{equation}
\sigma \Bigg\lbrace  {\frac{\partial v_{1}}{\partial \xi}} 
- {\frac{\partial \rho_{1}}{\partial \xi}}   \Bigg\rbrace+
\sigma^{2} \Bigg\lbrace {\frac{\partial v_{2}}{\partial \xi}}
-{\frac{\partial\rho_{2}}{\partial \xi}}
+{\frac{\partial\rho_{1}}{\partial \tau}}
+\rho_{1}{\frac{\partial v_{1}}{\partial \xi}}+v_{1}{\frac{\partial \rho_{1}}{\partial \xi}}
-{c_{s}}^{2}v_{1}{\frac{\partial v_{1}}{\partial \xi}} \Bigg\rbrace =0 
\label{contexp}
\end{equation}
respectively. In the last two equations each bracket must vanish independently and so
$\lbrace \dots  \rbrace = 0$. From  the first term of (\ref{contexp})  we obtain $\rho_{1}=v_{1}$. 
Using  this identity in the first term of (\ref{eulerexp}) we obtain an equation, which solved for
$c_s$ yields:
%proportional to $\sigma$ we obtain the constant $A$:
%\begin{equation}
%A \equiv \bigg({\frac{27g^{2}\,{\rho_{0}}^{2}}{{m_{G}}^{2}}}\bigg){c_{s}}^{2}
%+3\pi^{2/3}{\rho_{0}}^{4/3}{c_{s}}^{2}=\bigg({\frac{27g^{2}\,{\rho_{0}}^{2}}{{m_{G}}^{2}}}\bigg)
%+\pi^{2/3}{\rho_{0}}^{4/3}
%\label{acons}
%\end{equation}
%which gives the expression of the speed of sound calculation for a given reference 
%density ${\rho_{0}}$:
\begin{equation}
{c_{s}}^{2}={\frac{\bigg({\frac{27g^{2}\,{\rho_{0}}^{2}}{{m_{G}}^{2}}}\bigg)
+\pi^{2/3}{\rho_{0}}^{4/3}}
{\bigg({\frac{27g^{2}\,{\rho_{0}}^{2}}{{m_{G}}^{2}}}\bigg)+3\pi^{2/3}{\rho_{0}}^{4/3}}}
\label{csrelation}
\end{equation}
Inserting these results into the terms proportional to $\sigma^{2}$, we find, after some algebra 
the KdV equation:
$$
{\frac{\partial\rho_{1}}{\partial \tau}}
+\bigg[{\frac{(2-{c_{s}}^{2})}{2}}-\bigg({\frac{27g^{2}\,{\rho_{0}}^{2}}{{m_{G}}^{2}}}\bigg)
{\frac{(2{c_{s}}^{2}-1)}{2A}}-{\frac{\pi^{2/3}{\rho_{0}}^{4/3}}{A}}\bigg({c_{s}}^{2}
-{\frac{1}{6}}\bigg)\bigg]
\rho_{1}{\frac{\partial \rho_{1}}{\partial \xi}}
$$
\begin{equation}
+\bigg[{\frac{9g^{2}\,{\rho_{0}}^{2}}{{m_{G}}^{4}R^{2}A}}\bigg]
{\frac{\partial^{3} \rho_{1}}{\partial \xi^{3}}}=0  
\label{kdvxitau}
\end{equation}
where:
\begin{equation}
A = \bigg({\frac{27g^{2}\,{\rho_{0}}^{2}}{{m_{G}}^{2}}}\bigg){c_{s}}^{2}
+3\pi^{2/3}{\rho_{0}}^{4/3}{c_{s}}^{2}=\bigg({\frac{27g^{2}\,{\rho_{0}}^{2}}{{m_{G}}^{2}}}\bigg)
+\pi^{2/3}{\rho_{0}}^{4/3}
\label{acons}
\end{equation}
Returning to the  $x-t$ space we obtain:
\begin{equation}
{\frac{\partial \hat\rho_{1}}{\partial t}}+c_{s}{\frac{\partial \hat\rho_{1}}{\partial x}}
+\alpha {c_{s}} \hat\rho_{1}{\frac{\partial \hat\rho_{1}}{\partial x}}
+\beta{\frac{\partial^{3} \hat\rho_{1}}{\partial x^{3}}}=0
\label{kdvqcdg}
\end{equation}
where
\begin{equation}
\alpha \equiv \bigg[{\frac{(2-{c_{s}}^{2})}{2}}
-\bigg({\frac{27g^{2}\,{\rho_{0}}^{2}}{{m_{G}}^{2}}}\bigg)
{\frac{(2{c_{s}}^{2}-1)}{2A}}-{\frac{\pi^{2/3}{\rho_{0}}^{4/3}}{A}}\bigg({c_{s}}^{2}
-{\frac{1}{6}}\bigg)\bigg]
\label{alfa}
\end{equation}
and
\begin{equation}
\beta = \bigg[{\frac{9g^{2}\,{\rho_{0}}^{2} {c_{s}}}{{m_{G}}^{4}A}}\bigg]
\label{beta}
\end{equation}
%$$
%{\frac{\partial\hat\rho_{1}}{\partial t}}
%+{c_{s}}{\frac{\partial\hat\rho_{1}}{\partial x}}+
%\bigg[{\frac{(2-{c_{s}}^{2})}{2}}-\bigg({\frac{27g^{2}\,{\rho_{0}}^{2}}{{m_{G}}^{2}}}\bigg)
%{\frac{(2{c_{s}}^{2}-1)}{2A}}
%-{\frac{\pi^{2/3}{\rho_{0}}^{4/3}}{A}}\bigg({c_{s}}^{2}-{\frac{1}{6}}\bigg)\bigg]{c_{s}}
%\hat\rho_{1}{\frac{\partial\hat\rho_{1}}{\partial x}}
%$$
%\begin{equation}
%+\bigg[{\frac{27g^{2}\,{\rho_{0}}^{2}(1-{c_{s}}^{2}){c_{s}}}{4{m_{G}}^{4}A}}\bigg]
%{\frac{\partial^{3}\hat\rho_{1}}{\partial x^{3}}}=0  
%\label{kdvxt}
%\end{equation}
which is the KdV equation at zero temperature for the small perturbation in the  
baryon density
$\hat\rho_{1}\equiv \sigma\rho_{1}$.  
%Taking the limit $m_{G} \, \rightarrow \, \infty$ \, or \, $g=0$ \, we obtain 
%from (\ref{csrelation}):
%$$
%{c_{s}}^{2}={\frac{1}{3}}
%$$
%(\ref{kdvqcdg}) becomes:
%\begin{equation}
%{\frac{\partial\hat\rho_{1}}{\partial t}}+ 
%c_{s}{\frac{\partial \hat\rho_{1}}{\partial x}}+
%{\frac{2}{3}}c_{s}\hat\rho_{1}{\frac{\partial \hat\rho_{1}}{\partial x}}=0 
%\label{bwmitxitauXt}
%\end{equation}
%and we recover the  breaking wave equation (BW) for $\hat\rho_{1}$ in the QGP  
%obtained in \cite{fn5} from the MIT equation of state.  
If we neglect  the space and time derivatives in (\ref{eps}) and (\ref{pres}) and repeat 
the derivation sketched above we arrive at:
\begin{equation}
{\frac{\partial\hat\rho_{1}}{\partial t}}
+{c_{s}}{\frac{\partial\hat\rho_{1}}{\partial x}} +
%\bigg[{\frac{(2-{c_{s}}^{2})}{2}}-\bigg({\frac{27g^{2}\,{\rho_{0}}^{2}}{{m_{G}}^{2}}}\bigg)
%{\frac{(2{c_{s}}^{2}-1)}{2A}}-{\frac{\pi^{2/3}{\rho_{0}}^{4/3}}{A}}\bigg({c_{s}}^{2}
%-{\frac{1}{6}}\bigg)\bigg] 
 \alpha {c_{s}} \hat\rho_{1}{\frac{\partial\hat\rho_{1}}{\partial x}}=0  
\label{bwqxt}
\end{equation}
which is  a breaking wave (BW) equation for $\hat\rho_{1}$. We close this section 
emphasizing that Eq. (\ref{kdvqcdg}) is the main result of this work. It shows that for 
suitable choices of the parameters $\alpha$ and $\beta$ we can have KdV solitons in a quark 
gluon plasma.

\section{Numerical analysis }

The KdV equation (\ref{kdvqcdg}) has an analytical soliton solution given by \cite{drazin}:
\begin{equation}
\hat\rho_{1}(x,t)={\frac{3(u-c_{s})}{\alpha c_s}} \ sech^{2}
\Bigg[{\sqrt{{\frac{(u-c_{s})}{4\beta}}}}(x-ut)\Bigg]
\label{soliton}
\end{equation}
where $u$ is an arbitrary supersonic velocity.
A soliton is a localized pulse which propagates without change in shape and in this 
case it has the width $\lambda$
defined by:
\begin{equation}
\lambda={\sqrt{{\frac{4\beta}{(u-c_{s})}}}}=
{\sqrt{{\frac{36g^{2}\,{\rho_{0}}^{2}{c_{s}}}{(u-c_{s}){m_{G}}^{4}A}}}}
\label{solitonwidth}
\end{equation}
For the numerical estimates we shall use the following values of the parameters : 
$\mathcal{B}_{QCD}= \, 0.0006 \,\, \mbox{GeV}^{4}$, $g = 0.35$, 
$m_G = 290 $ MeV and $\rho_{0}=2 \, fm^{-3}$ 
which, when sustituted  in    (\ref{csrelation}) yield ${c_{s}}^{2}=0.5$ \, $(c_{s}=0.7)$.
We choose $u=0.8$. With the help of  (\ref{solitonwidth}) we find  
$\lambda= 1.7 \, fm$ \, for the soliton width   and  $0.6$ for its amplitude. 
Even though this work is essentially qualitative 
the chosen numbers are well appropriate to study a realistic situation of a perturbation 
traversing the QGP.

In Fig. 1 we show the numerical solution of (\ref{kdvqcdg}) with initial condition  
$\hat\rho_{1}(x,t=0)$ and $\hat\rho_{1}(x,t)$ given by  (\ref{soliton}).  
We can  observe the time evolution of the initial 
gaussian-like pulse as a well defined soliton, keeping its shape and form. This solution 
shows the behavior expected from the analytic solution. We see that using the correct input 
for the amplitude and width we obtain a pulse which propagates without distortion.  
This initial 
condition  is a very special case of little practical interest. A real perturbation 
produced in the 
QGP will most likely have the ``wrong'' amplitude and ``wrong'' width. For arbitrary amplitudes 
the solution must be numerically calculated.

In Fig. 2 we show again the numerical solution of (\ref{kdvqcdg}) with the initial condition given by   
(\ref{soliton}) multiplied by a factor $10$. Now we observe that the initial pulse starts to develop 
secondary peaks, which are called ``radiation'' in the literature.  Further time evolution will increase 
the  strength of these peaks until the complete loss of localization. 

In Fig. 3 we show the numerical solution   of  (\ref{bwqxt}) with the initial condition given by 
(\ref{soliton}).  We observe the gradual formation of a ``wall''  followed by the dispersion of the initial 
pulse. In Fig. 4  the amplitude of the initial pulse  (\ref{soliton}) is multiplied by a factor $10$ and 
used as initial condition for (\ref{bwqxt}).  As expected the dispersion takes place much earlier than in 
Fig. 3.

\begin{figure}[h]
\begin{center}
\epsfig{file=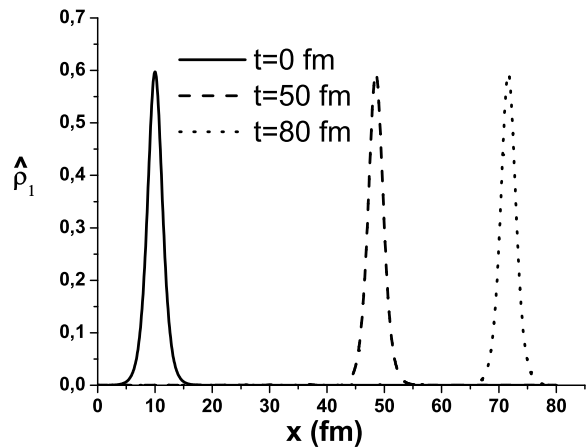,width=86mm}
\caption{Numerical solution of (\ref{kdvqcdg}) with (\ref{soliton}) as initial condition 
calculated at different times. }
\end{center}
\label{fig1}
\end{figure}

\begin{figure}[h]
\begin{center}
\epsfig{file=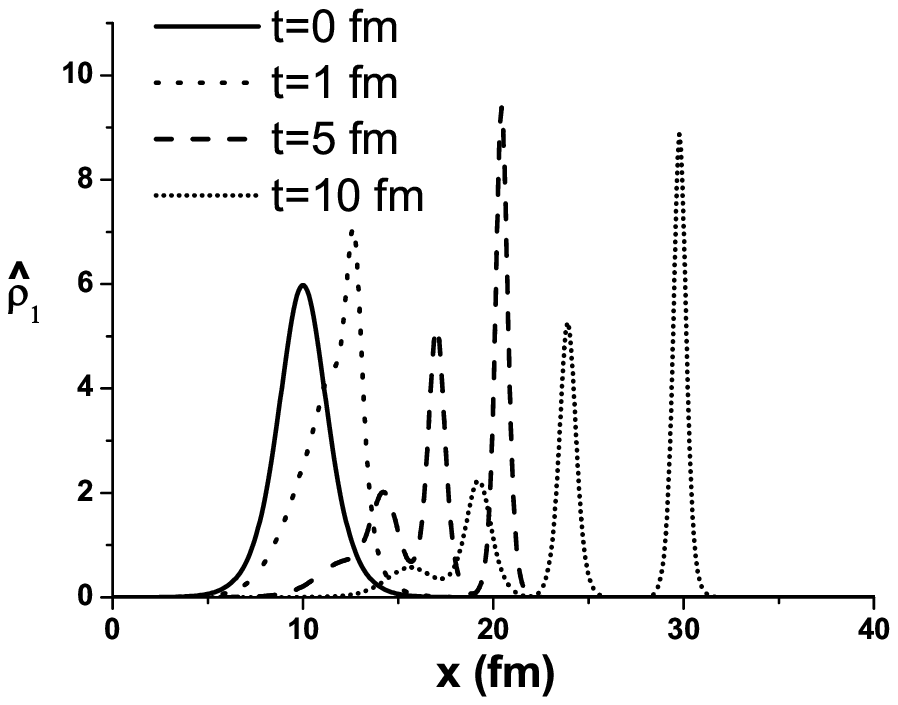,width=86mm}
\caption{The same as Fig. 1 with an initial amplitude ten times larger.}
\end{center}
\label{fig2}
\end{figure}

\begin{figure}[h]
\begin{center}
\epsfig{file=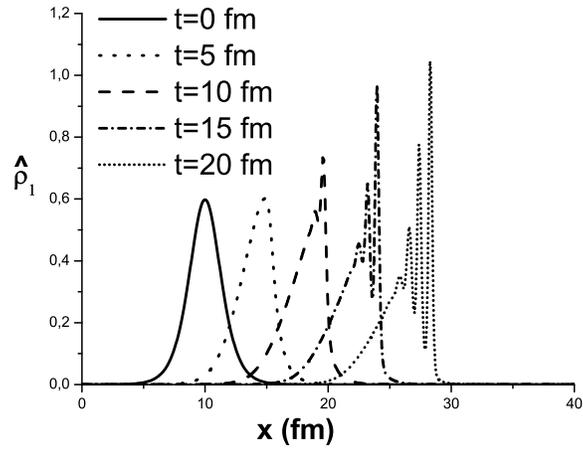,width=86mm}
\caption{Numerical solution of (\ref{bwqxt}) with (\ref{soliton}) as initial condition. }
\end{center}
\label{fig3}
\end{figure}

\begin{figure}[h]
\begin{center}
\epsfig{file=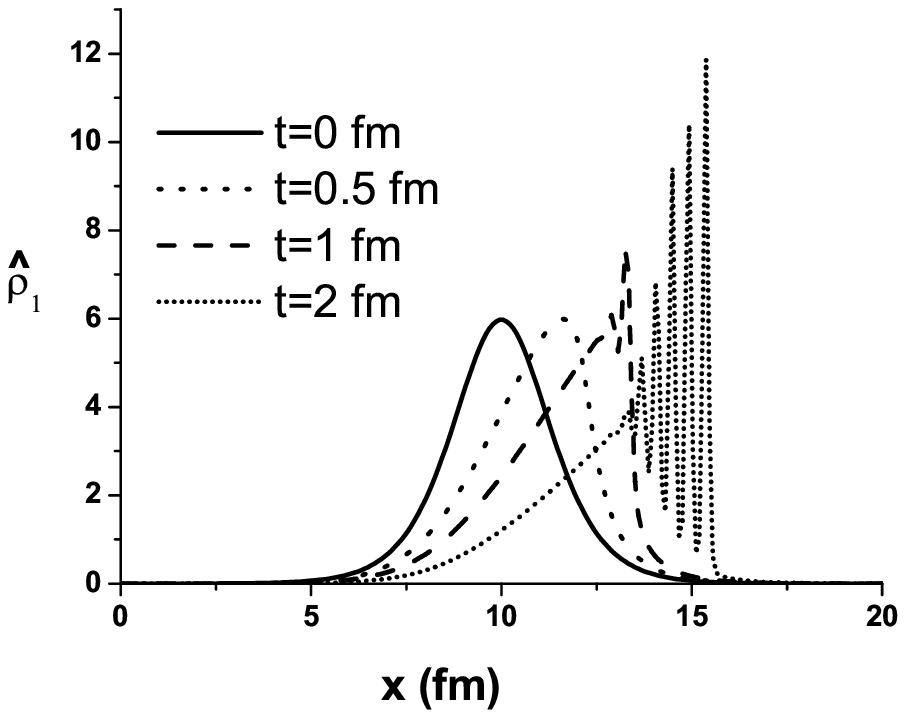,width=86mm}
\caption{The same as Fig. 3 with an initial amplitude ten times larger.}
\end{center}
\label{fig4}
\end{figure}

\section{Summary}

The main conclusion of this work is that it is indeed possible to have KdV solitons in QCD, 
provided that two conditions are satisified.  
The first condition is that the gluon field have a dynamical mass. In this case the equation 
of motion  (\ref{azeaem}) can be solved in the weak inhomogeneity approximation yielding 
(\ref{azerototale}).  The existence of a dynamical gluon mass has been intensely discussed in 
the literature  during the last years  and seems to be well established (for details 
see  the references given in \cite{fn6}).  For a massless gluon 
field  we can only have a breaking wave equation, as it was found in our previous 
work \cite{fn5}. 
The second condition  is the  existence of second order derivative terms  in
the energy density and pressure.  These terms appear naturally from the formalism, as we can
see in (\ref{epsilonnnn}). However  it is necessary to keep these derivative terms. The
use of uniform field approximations  prevents us from finding KdV solitons. If we neglect the 
derivatives we arrive at the  breaking wave equation (\ref{bwqxt}).  The  practical difference 
between perturbations governed by the KdV and BW equations is that the former propagate much 
longer keeping its localization whereas the latter loose localization and  may generate 
unstable ``walls''. The numerical analysis of some cases  confirms the anticipated qualitative 
expectation. The application of the formalism developed in this work to problems in the theory 
of compact stars is in progress.

\begin{acknowledgments}
%We are deeply grateful to S. Raha for useful discussions.
This work was  partially financed by the Brazilian funding
agencies CAPES, CNPq and FAPESP. 
\end{acknowledgments}


\begin{thebibliography}{99}
				  
\bibitem{drazin}  P. G. Drazin and R. S. Johnson, ``Solitons: An Introduction'', Cambridge 
University Press, 1989.			


\bibitem{frsw} G.N. Fowler, S. Raha, N. Stelte and R.M. Weiner, 
               Phys. Lett. B {\bf 115}, 286 (1982); 
               S. Raha, K. Wehrberger and R.M. Weiner, 
               Nucl. Phys. A {\bf 433}, 427 (1984);  
               E.F. Hefter, S. Raha and R.M. Weiner, 
               Phys. Rev.  C {\bf 32}, 2201 (1985).

%\bibitem{abu} A.Y. Abul-Magd, I. El-Taher and F.M. Khaliel, 
%              Phys. Rev. C {\bf 45}, 448 (1992). 
 
\bibitem{fn1} D.A. Foga\c{c}a and  F.S. Navarra, Phys. Lett. B {\bf 639}, 629 (2006).
 
\bibitem{fn2} D.A. Foga\c{c}a and  F.S. Navarra, Phys. Lett. B {\bf 645}, 408 (2007). 

\bibitem{fn3} D.A. Foga\c{c}a and  F.S. Navarra,  Nucl. Phys. A {\bf 790}, 619c (2007); 
              Int. J. Mod. Phys. E {\bf 16}, 3019 (2007). 
	
\bibitem{fn4} D.A. Foga\c{c}a, L. G. Ferreira Filho and  F.S. Navarra, 
              Nucl. Phys. A {\bf 819}, 150 (2009).

\bibitem{fn5} D.A. Foga\c{c}a, L.G. Ferreira Filho and  F.S. Navarra,  
              Phys. Rev. C {\bf 81}, 055211 (2010).		

\bibitem{fn6} D.A. Foga\c{c}a and  F.S. Navarra, 
              Phys. Lett. B {\bf 700}, 236 (2011).		  

\bibitem{wein} S. Weinberg,``Gravitation and Cosmology'', New York: Wiley, 1972.


\bibitem{land} L. Landau and  E. Lifchitz, ``Fluid Mechanics'',  
                Pergamon Press, Oxford, (1987).				    

\bibitem{shakin} L. S. Celenza and C. M. Shakin, Phys.  Rev. D {\bf 34}, 1591 (1986).

\bibitem{shakinn} X. Li and C. M. Shakin, Phys.  Rev. D {\bf 71}, 074007 (2005).

\bibitem{serot} B.D. Serot and J.D. Walecka, 
                 Advances in Nuclear Physics {\bf 16}, 1 (1986).
			   				 				  
\bibitem{davidson} R. C. Davidson, ``Methods in Nonlinear Plasma Theory'', 
                   Academic Press, 
                   New York an London, 1972, pages 20 and 21.				

				 	 			   
\end{thebibliography}
\end{document}